\documentclass[twocolumn,aps,pra,showpacs,groupedaddress,superscriptaddress,nofootinbib]{revtex4-1}

\usepackage[pdftex,linkcolor=blue,citecolor=blue,urlcolor=blue,colorlinks]{hyperref}
\usepackage[dvipsnames]{xcolor}
\usepackage{graphicx,epsfig}
\usepackage{color}
\usepackage{cancel}
\usepackage{inputenc}
\usepackage{amsmath}
\usepackage{amsfonts}
\usepackage{fancyhdr}
\usepackage{ulem}

\newcommand{\ket}[1]{\ensuremath{\left| #1 \right\rangle}}
\newcommand{\bra}[1]{\ensuremath{\left\langle #1 \right|}}
\newcommand{\braket}[2]{\ensuremath{\left\langle #1 | #2\right\rangle}}
\newcommand{\brakett}[3]{\ensuremath{\left\langle #1 \left| #2\right| #3\right\rangle}}

\newcommand{\refsec}[1]{Sec.~\ref{#1}}
\newcommand{\refeq}[1]{Eq.~(\ref{#1})}
\newcommand{\refeqs}[1]{Eqs.~(\ref{#1})}

\newcommand{\1}{\ensuremath{\left|1\right\rangle}}

\newcommand{\fref}[1]{Fig.~\ref{#1}}

\newcommand{\cref}[1]{chapter~\ref{#1}}
\newcommand{\Cref}[1]{Chapter~\ref{#1}}

\newcommand{\figref}[1]{Fig.~\ref{#1}}

\begin{document}
\title{Interaction-induced topological properties of two bosons in flat-band systems}
\author{G. Pelegr\'i}
\affiliation{Departament de F\'isica, Universitat Aut\`onoma de Barcelona, E-08193 Bellaterra, Spain}
\affiliation{Department of Physics and SUPA, University of Strathclyde, Glasgow G4 0NG, UK}
\author{A. M. Marques}
\affiliation{Department of Physics and I3N, University of Aveiro, 3810-193 Aveiro, Portugal}
\author{V. Ahufinger}
\affiliation{Departament de F\'isica, Universitat Aut\`onoma de Barcelona, E-08193 Bellaterra, Spain}
\author{J. Mompart}
\affiliation{Departament de F\'isica, Universitat Aut\`onoma de Barcelona, E-08193 Bellaterra, Spain}
\author{R. G. Dias}
\affiliation{Department of Physics and I3N, University of Aveiro, 3810-193 Aveiro, Portugal}

\begin{abstract}
In flat-band systems, destructive interference leads to the localization of non-interacting particles and forbids their motion through the lattice. However, in the presence of interactions the overlap between neighbouring single-particle localized eigenstates may enable the propagation of bound pairs of particles. In this work, we show how these interaction-induced hoppings can be tuned to obtain a variety of two-body topological states. In particular, we consider two interacting bosons loaded into the orbital angular momentum $l=1$ states of a diamond-chain lattice, wherein an effective $\pi$ flux may yield a completely flat single-particle energy landscape. In the weakly-interacting limit, we derive effective single-particle models for the two-boson quasiparticles which provide an intuitive picture of how the topological states arise. By means of exact diagonalization calculations, we benchmark these states and we show that they are also present for strong interactions and away from the strict flat-band limit. Furthermore, we identify a set of doubly localized two-boson flat-band states that give rise to a special instance of Aharonov-Bohm cages for arbitrary interactions.
\end{abstract}
\maketitle


\section{Introduction}
The study of topological materials is a prominent topic in condensed matter physics. One of the main features of these exotic systems is the existence of a bulk-boundary correspondence, which correlates the non-trivial topological indices of the bulk energy bands with the presence of robust edge modes \cite{TopInsReview}. At the single-particle level, topological phases are well understood and can be systematically classified in terms of their symmetries and dimensionality \cite{reviewTopIns2,TopInsSymReview}. However, many questions regarding the characterization of interacting topological systems remain open. While it is known that interactions are responsible for the appearence of strongly correlated many-body topological phases such as fractional quantum Hall states \cite{QuantumHallFrac} or symmetry-protected topological phases \cite{SPTSenthil}, topological invariants are in general difficult to define even in few-body systems, making the systematic study of interaction-driven topological effects ellusive.    

Over the last years, a number of topological models have been successfully implemented with ultracold atoms in optical lattices \cite{ReviewUltracoldTop} and photonic systems \cite{ReviewTopologicalPhotonics}. Due to their high degree of tunability and control, these synthetic platforms offer exciting perspectives for exploring interacting topological phases \cite{reviewColdTop2}. A remarkable example of this is the possibility to realize long-lived bound pairs of interacting particles, also known as doublons. These two-body states, which are stable even for repulsive interactions due to the finite bandwidth of the single-particle kinetic energy \cite{FewBodyLattice}, have been observed \cite{DoublonOriginal,DoublonDecay,DoublonQuantumWalk,DoublonHarperHofstadter} and extensively analyzed \cite{TwoBody1DLattice1,TwoBody1DLattice2,TwoBody1DLattice3,EdgeLocalized1DLattice1,TwoBody1DLattice4,TwoBody1DLattice5,
TwoBody1DLattice6,TwoBody1DLattice7,TwoBody1DLattice8,EdgeLocalized1DLattice2,TwoBody1DLattice9,TwoBody1DLattice10} in optical lattices, and have also been emulated in photonic systems \cite{TwoBody1DLatticePhotonics1,TwoBody1DLatticePhotonics2} and in topolectrical circuits \cite{TwoBodyTopologyTopolectrical}. Motivated in part by these advances, several recent works have focused on the topological properties of two-body states \cite{DoublonHarperHofstadter,SSH2bosons1,SSH2bosons2,SSHHolesChain1,SSHHolesChain2,SSH2particlesRing,SSH2bosons3,
Haldane2bosons,Topology2photons1,Topology2photons2,TopologyFermionicDoublons1,
TopologyFermionicDoublons2,DoublonLattices,TwoBodyCreutz,Ke-2017,Lin-2020,Lyubarov-2019}, with the long-term aim of paving the path to a better comprehension of topological phases in a full many-body interacting scenario. A distinctive advantage that these small-sized systems offer is that it is often possible to map the problem of two interacting particles in a lattice into a single-particle model defined in a different lattice, the topological characterization of which can then be performed with well-established techniques \cite{SSH2particlesRing,SSHHolesChain1,SSHHolesChain2,Topology2photons1,DoublonLattices,TwoBodyCreutz}.

In this paper, we study the topological properties of two-boson states in a system that is topologically non-trivial at the single-particle level. Specifically, we consider a diamond-chain lattice filled with ultracold atoms loaded into Orbital Angular Momentum (OAM) $l=1$ states. In this geometry, the OAM degree of freedom induces an effective $\pi$-flux which yields a single-particle spectrum composed entirely of flat bands upon a proper tuning of the tunneling parameters \cite{diamondchain1,diamondchain2}. In this situation, quantum interference leads to a strong localization of non-interacting particles and forbids their propagation through the chain. This phenomenon, known as Aharonov-Bohm caging \cite{ABcageoriginal,diamondchain1,diamondchain2,recentwork1,recentwork2, ABcagingPhotonics}, has been recently generalized to non-Abelian systems, where it is expected to yield intriguing state-dependent dynamics \cite{NonAbelianABcaging}. This single-particle localization effect is a general characteristic of flat-band systems \cite{ReviewFlatBands}, wherein the role of the kinetic energy becomes irrelevant and particle motion can only originate from interaction-mediated collective processes. This peculiar feature is responsible for the appearance in such systems of a number of exotic quantum states determined solely by the interactions and the geometry of the lattice \cite{DiamondChainManyBody,FlatBands1,FlatBands2,FlatBands3,FlatBands4,FlatBands5,FlatBands6,FlatBands7,FlatBands8,
FlatBands9,FlatBands10,FlatBands11,Kuno-2020}, including topologically non-trivial phases \cite{TopoFlatBands1,TopoFlatBands2,TopoFlatBands3,TopoFlatBands4,similar2}.

In this work, we focus on the limit of weak attractive interactions, in which the low-energy properties of the system can be studied by projecting the Hamiltonian into the lowest flat bands \cite{FlatBands1,FlatBands2,FlatBands3,FlatBands4}. By mapping the subspaces of lowest-energy two-boson states into single-particle models, we show that the system has a topologically non-trivial phase. In contrast with other realizations of two-body topological states  \cite{DoublonHarperHofstadter,SSH2bosons1,SSH2bosons2,SSHHolesChain1,SSHHolesChain2,SSH2particlesRing,SSH2bosons3,
Haldane2bosons,Topology2photons1,Topology2photons2,TopologyFermionicDoublons1,
TopologyFermionicDoublons2,DoublonLattices,TwoBodyCreutz}, in this case the topological character is controlled through effective two-boson tunneling amplitudes that depend on the interaction strength. In a diamond chain with open boundaries, this topological phase is benchmarked by the presence of robust in-gap states localized at the edges, which are in turn composed of bound pairs of bosons, each occupying a localized single-particle eigenstate. We provide numerical evidence that these edge states are also present for interaction strengths capable of introducing mixing with higher bands and in the case where the bands are not completely flat. Moreover, we find that the system displays doubly-localized flat bands of two-boson states that give rise to Ahoronov-Bohm caging also for interacting particles prepared in specific states. 

The rest of the paper is organized as follows. In \refsec{PhysicalSystem} we discuss the single-particle properties of the system and we provide a qualitative description of the effect of attractive on-site interactions on the properties of two-boson states in the flat-band limit. In \refsec{EffectiveModels} we derive effective tight-binding models for the lowest-energy sector of the two-boson spectrum by projecting the interacting part of the Hamiltonian into the lowest flat bands. We find that these effective models can be rendered topologically non-trivial by selectively tuning the strength of the interactions on the different sites of the lattice. We also use the models to identify the doubly-localized two-boson flat-band states. In \refsec{EDsec} we provide exact diagonalization results that confirm the analytical predictions of the previous section. We also examine numerically the robustness of the two-boson topological phases upon deviations from the weakly-interacting and flat-band conditions. In \refsec{Experiment} we address some aspects of the possible experimental realization. Finally, in \refsec{Conclusions} we summarize the main conclusions of this work.  
\section{Model}
\label{PhysicalSystem}
\begin{figure}[t!]
\includegraphics[width=\linewidth]{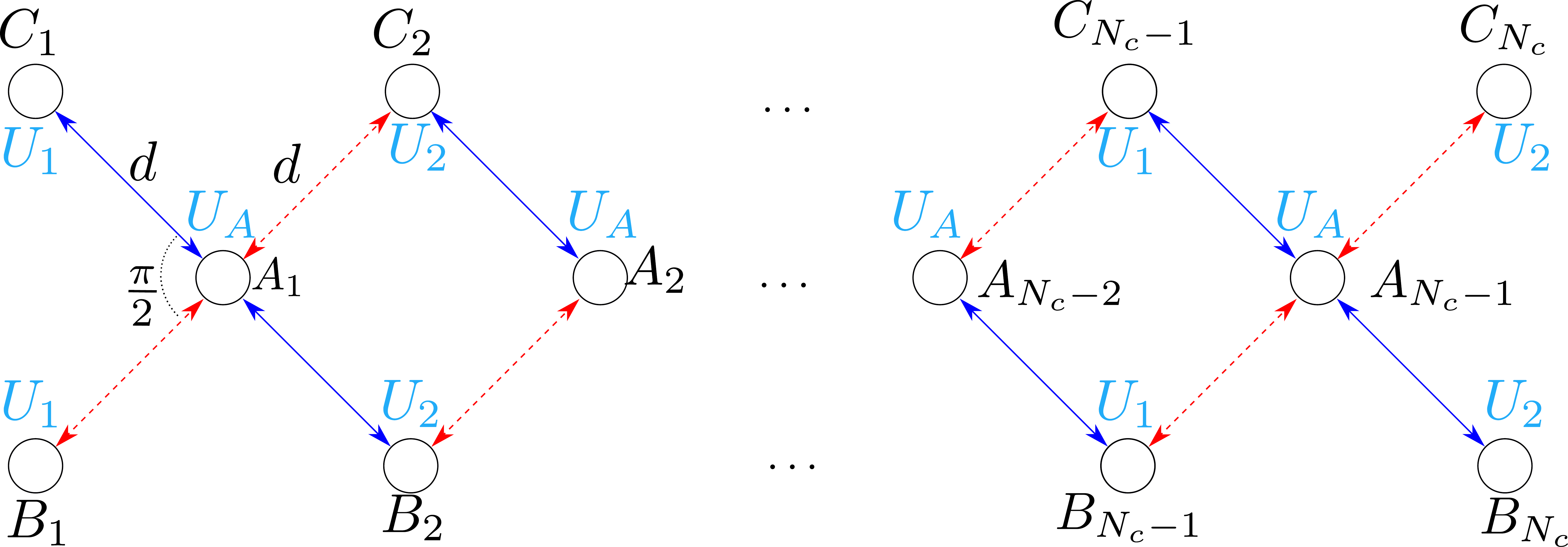}
\caption{Schematic representation of the diamond-chain optical lattice considered in this work. The $J_3$ couplings of \refeq{Hkin_dc} are real and positive along the directions marked by solid blue arrows, whereas they acquire a relative $\pi$ phase along the directions marked by dotted red arrows. The $J_2$ are real and positive along all directions. The on-site interaction strength is $U_A$ at the $A$ sites of each unit cell and $U_1$ ($U_2$) at the $B$ and $C$ sites of odd (even) unit cells.}
\label{PhySystem}
\end{figure}

We consider a gas of cold bosonic atoms trapped in a diamond-chain optical lattice. As shown in \fref{PhySystem}, the unit cells of the lattice, which we label with the index $i$, consist of a spinal site $A_i$ and two sites $B_i$ and $C_i$ equally separated from $A_i$ by a distance $d$ and forming a relative angle $\Theta=\pi/2$. Each of the sites corresponds to the center of a cylindrically symmetric potential of radial frequency $\omega$, which naturally supports OAM states. The atoms are loaded into the manifold formed by the two degenerate OAM $l=1$ states with positive and negative circulation localized at each site, $\ket{j_i,\pm}$, the wavefunctions of which are given by
\begin{equation}
\phi_{\alpha}^{j_i}(r_{j_i},\varphi_{j_i})=\braket{\vec{r}}{j_i,\pm}=\psi(r_{j_i})e^{\pm i(\varphi_{j_i}-\varphi_0)},
\label{wavefunctions}
\end{equation}
where $j\in \{A,B,C\}$, $(r_{j_i},\varphi_{j_i})$ are the polar coordinates with origin at site $j_i$, and $\varphi_0$ is an arbitrary phase origin.
Let us first briefly recall the non-interacting description of this system, which was analyzed in detail in \cite{diamondchain1,diamondchain2}. In general, the single-particle dynamics of ultracold atoms carrying OAM in arrays of side-coupled potentials are governed by three coupling amplitudes \cite{geometricallyinduced}, which we denote as $J_1$, $J_2$ and $J_3$. Specifically, $J_1$ accounts for the coupling between the two OAM states with opposite circulation within one site, while $J_2$ and $J_3$ correspond to tunneling processes between OAM states localized at neighbouring sites with equal and opposite circulations, respectively. For the particular case of the diamond-chain geometry, the self-coupling $J_1$ vanishes everywhere except at the edges of the chain, so its effect can be neglected \cite{diamondchain1}. Following the same criterion as in \cite{diamondchain1,diamondchain2}, we fix $\varphi_0$ along the line that connects the sites $C_i\leftrightarrow A_i \leftrightarrow B_{i+1}$, indicated with solid blue arrows in \figref{PhySystem}. With this choice of the phase origin, the $J_3$ amplitudes along the line connecting the sites $B_i\leftrightarrow A_i \leftrightarrow C_{i+1}$, indicated with dotted red arrows in \figref{PhySystem}, acquire a relative $\pi$ phase. Thus, taking $J_2$ and $J_3$ as positive quantities \cite{diamondchain1}, the non-interacting Hamiltonian of the system reads
\begin{align}
\hat{H}_0&=J_2\sum_{i=1}^{N_c-1}\sum_{\alpha=\pm}\left[\hat{a}_{\alpha}^{i\dagger}(\hat{b}_{\alpha}^{i}+\hat{b}_{\alpha}^{i+1}+\hat{c}_{\alpha}^{i}+\hat{c}_{\alpha}^{i+1})\right]\nonumber\\
&+J_3\sum_{i=1}^{N_c-1}\sum_{\alpha=\pm}\left[\hat{a}_{\alpha}^{i\dagger}(-\hat{b}_{-\alpha}^{i}+\hat{b}_{-\alpha}^{i+1}
+\hat{c}_{-\alpha}^{i}-\hat{c}_{-\alpha}^{i+1})\right]\nonumber\\
&+\text{H.c.},
\label{Hkin_dc}
\end{align}
where, $\hat{a}_{\alpha}^i$, $\hat{b}_{\alpha}^i$ and $\hat{c}_{\alpha}^i$ are the bosonic annihilation operators associated to the OAM states $\ket{A_i,\alpha}$, $\ket{B_i,\alpha}$ and $\ket{C_i,\alpha}$ respectively, and $N_c$ is the total number of unit cells forming the chain. For reasons that we explain below, we do not consider the $A$ site of the last unit cell of the chain.

\begin{figure}[h!]
\includegraphics[width=\linewidth]{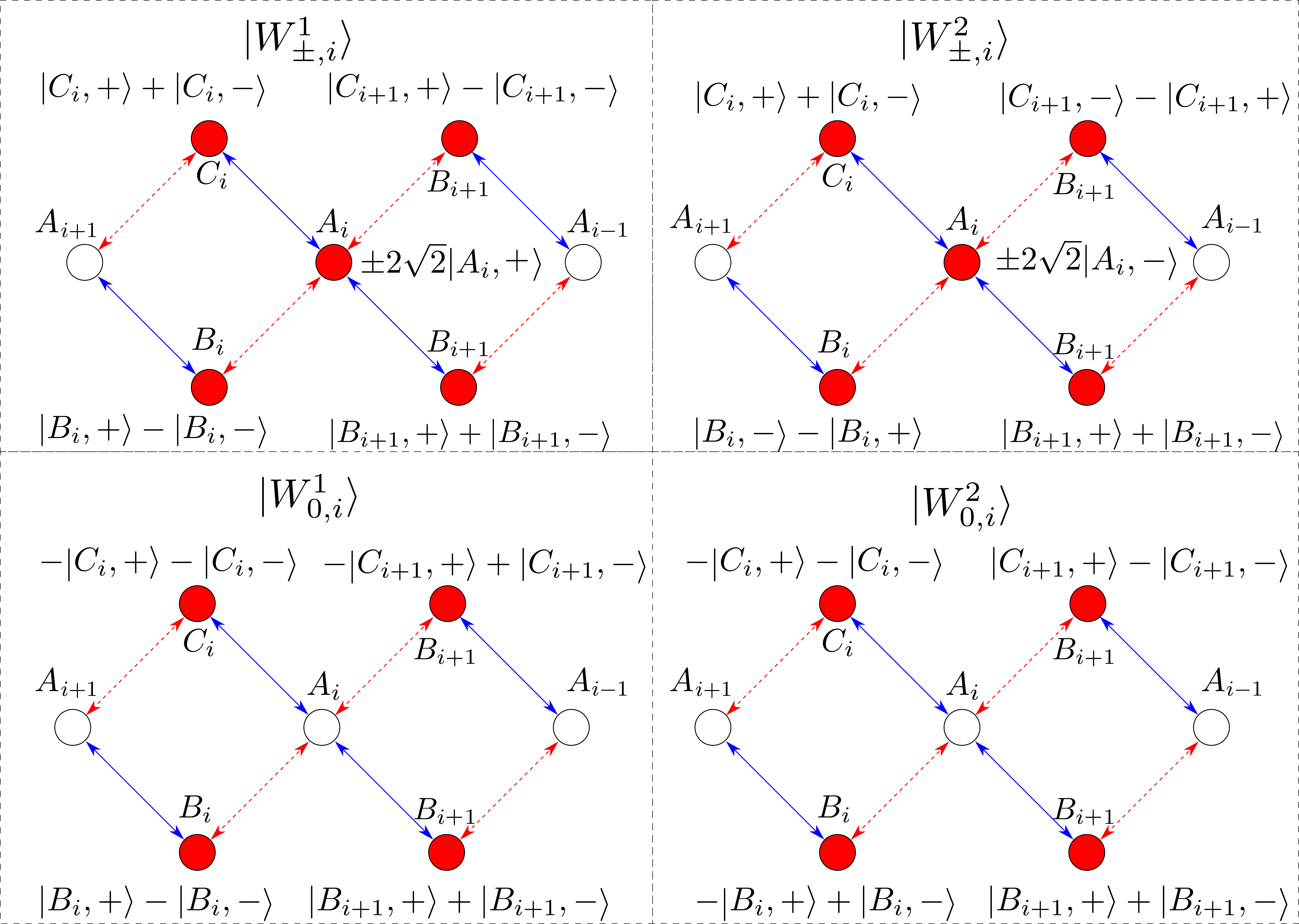}
\caption{Sketches of the maximally localized flat-band eigenstates created by the operators given in Eqs. \eqref{LowerBand}-\eqref{UpperBand} indicating the relative weights of each of the OAM states at the different sites.}
\label{FlatStates}
\end{figure}

In the present study we also consider the effect of on-site interactions between the bosons. As shown in \fref{PhySystem}, we let the interaction strength take the value $U_A$ in the $A$ sites of each unit cell and $U_1$ and $U_2$ in the $B$ and $C$ sites of odd and even unit cells, respectively. As we will explain in \refsec{EffectiveModels}, tuning independently these different interaction parameters allows to explore distinct topological regimes of two-boson states. Such modulation of the onsite interactions could be achieved either by creating a lattice with slightly different trapping frequencies at each site or by means of magnetic \cite{ReviewFeshbach} or optical \cite{OpticalFeshbach3} Feshbach resonances induced by a spatially modulated field. Assuming that the interactions occur only through low-energy collisions between bosons in OAM $l=1$ states occupying the same site, we can write the part of the Hamiltonian describing the interactions as \cite{spinmodel}
\begin{align}
\hat{H}_{\text{int}}&=\frac{U_A}{2}\sum_{i=1}^{N_c}\hat{n}_{+}^{a_i}(\hat{n}_{+}^{a_i}-1)
+\hat{n}_{-}^{a_i}(\hat{n}_{-}^{a_i}-1)+4\hat{n}_{+}^{a_i}\hat{n}_{-}^{a_i}\nonumber\\
&+\frac{U_1}{2}\sum_{i, \text{odd}}\sum_{j=b,c}\hat{n}_{+}^{j_{i}}(\hat{n}_{+}^{j_{i}}-1)
+\hat{n}_{-}^{j_{i}}(\hat{n}_{-}^{j_{i}}-1)+4\hat{n}_{+}^{j_{i}}\hat{n}_{-}^{j_{i}}\nonumber\\
&+\frac{U_2}{2}\sum_{i, \text{even}}\sum_{j=b,c}\hat{n}_{+}^{j_{i}}(\hat{n}_{+}^{j_{i}}-1)
+\hat{n}_{-}^{j_{i}}(\hat{n}_{-}^{j_{i}}-1)+4\hat{n}_{+}^{j_{i}}\hat{n}_{-}^{j_{i}},
\label{Hint_dc}
\end{align}
where we have defined the number operators $\hat{n}_{\pm}^{j_i}\equiv \hat{j}_{\pm}^{i\dagger}\hat{j}_{\pm}^{i}$. Finally, the total Hamiltonian reads
\begin{equation}
\hat{H}=\hat{H}_0+\hat{H}_{\text{int}}.
\label{HamTotal}
\end{equation}
Let us start by addressing the single-particle properties of the system. In a chain with periodic boundary conditions, the spectrum of the non-interacting Hamiltonian \eqref{Hkin_dc} is composed of three two-fold degenerate bands \cite{diamondchain1,diamondchain2}
\begin{subequations}
\begin{align}
&E_{-}^{1}(k)=E_{-}^{2}(k)=-2\sqrt{(J_2^2+J_3^2)+(J_2^2-J_3^2)\cos k}\\
&E_{0}^{1}(k)=E_{0}^{2}(k)=0,\\
&E_{+}^{1}(k)=E_{+}^{2}(k)=2\sqrt{(J_2^2+J_3^2)+(J_2^2-J_3^2)\cos k},
\end{align}
\end{subequations}
where we set the lattice spacing to $a=1$ here and throughout the paper. We focus on the $J_2=J_3$ limit, which can be approximately realized by setting a large value of the separation between sites $d$ \cite{diamondchain1, diamondchain2}. In this limit, all the bands become flat with energies $E_{-}^{1,2}=-2\sqrt{2}J_2,E_{0}^{1,2}=0,E_{+}^{1,2}=2\sqrt{2}J_2$. For these values of the tunneling parameters, a diamond chain with open boundaries also displays in-gap states with energies $E = \pm 2J_2$ localized at the right edge. Nevertheless, these states can be removed from the single-particle spectrum by cutting the $A$ site of the last unit cell, $A_{N_c}$, in such a way that the left and right ends of the chain are symmetric, as shown in \figref{PhySystem}. From now onwards, we will assume that there are no single-particle edge states of this kind in the lattice. The eigenstates belonging to each of the flat bands can be expressed as compact modes that are completely localized in two consecutive unit cells. Specifically, the bosonic creation operators associated with the flat-band modes localized in the $i$ and $i+1$ unit cells can be written in terms of the original OAM creation operators as \cite{diamondchain1,diamondchain2}
\begin{subequations}
\begin{align}
\hat{W}_{-,i}^{1\dagger}=\frac{1}{4}&\left(-\frac{4}{\sqrt{2}}\hat{a}_{+}^{i\dagger}+\hat{b}_{+}^{i\dagger}-\hat{b}_{-}^{i\dagger}+\hat{c}_{+}^{i\dagger}+\hat{c}_{-}^{i\dagger}\right.\nonumber\\
&\left.+\hat{b}_{+}^{i+1\dagger}+\hat{b}_{-}^{i+1\dagger}+\hat{c}_{+}^{i+1\dagger}-\hat{c}_{-}^{i+1\dagger}\right),\\
\hat{W}_{-,i}^{2\dagger}=\frac{1}{4}&\left(-\frac{4}{\sqrt{2}}\hat{a}_{-}^{i\dagger}-\hat{b}_{+}^{i\dagger}+\hat{b}_{-}^{i\dagger}+\hat{c}_{+}^{i\dagger}+\hat{c}_{-}^{i\dagger}\right.\nonumber\\
&\left.+\hat{b}_{+}^{i+1\dagger}+\hat{b}_{-}^{i+1\dagger}
-\hat{c}_{+}^{i+1\dagger}+\hat{c}_{-}^{i+1\dagger}\right),
\end{align}
\label{LowerBand}
\end{subequations}
for the states in the lower bands, leading to 
\begin{equation}
\hat{H}_{0}(J_2=J_3)\left(\hat{W}_{-,i}^{1/2\dagger}\ket{0}\right)=
-2\sqrt{2}J_2\left(\hat{W}_{-,i}^{1/2\dagger}\ket{0}\right).    
\end{equation}
For the states in the middle bands we have
\begin{subequations}
\begin{align}
\hat{W}_{0,i}^{1\dagger}=\frac{1}{2\sqrt{2}}&\left(-\hat{b}_{+}^{i\dagger}+\hat{b}_{-}^{i\dagger}+\hat{c}_{+}^{i\dagger}+\hat{c}_{-}^{i\dagger}
\right.\nonumber\\
&\left.-\hat{b}_{+}^{i+1\dagger}-\hat{b}_{-}^{i+1\dagger}+\hat{c}_{+}^{i+1\dagger}-\hat{c}_{-}^{i+1\dagger}\right),
\\
\hat{W}_{0,i}^{2\dagger}=\frac{1}{2\sqrt{2}}&\left(\hat{b}_{+}^{i\dagger}-\hat{b}_{-}^{i\dagger}+\hat{c}_{+}^{i\dagger}+\hat{c}_{-}^{i\dagger}
\right.\nonumber\\
&\left.-\hat{b}_{+}^{i+1\dagger}-\hat{b}_{-}^{i+1\dagger}
-\hat{c}_{+}^{i+1\dagger}+\hat{c}_{-}^{i+1\dagger}\right),
\end{align}
\label{ZeroBand}
\end{subequations}
leading to
\begin{equation}
\hat{H}_{0}(J_2=J_3)\left(\hat{W}_{0,i}^{1/2\dagger}\ket{0}\right)=0.    
\end{equation}
Finally, for the states in the upper bands we have
\begin{subequations}
\begin{align}
\hat{W}_{+,i}^{1\dagger}=\frac{1}{4}&\left(\frac{4}{\sqrt{2}}\hat{a}_{+}^{i\dagger}+\hat{b}_{+}^{i\dagger}-\hat{b}_{-}^{i\dagger}+\hat{c}_{+}^{i\dagger}+\hat{c}_{-}^{i\dagger}\right.\nonumber\\
&\left.+\hat{b}_{+}^{i+1\dagger}+\hat{b}_{-}^{i+1\dagger}+\hat{c}_{+}^{i+1\dagger}-\hat{c}_{-}^{i+1\dagger}\right),\\
\hat{W}_{+,i}^{2\dagger}=\frac{1}{4}&\left(\frac{4}{\sqrt{2}}\hat{a}_{-}^{i\dagger}-\hat{b}_{+}^{i\dagger}+\hat{b}_{-}^{i\dagger}+\hat{c}_{+}^{i\dagger}+\hat{c}_{-}^{i\dagger}\right.\nonumber\\
&\left.+\hat{b}_{+}^{i+1\dagger}+\hat{b}_{-}^{i+1\dagger}
-\hat{c}_{+}^{i+1\dagger}+\hat{c}_{-}^{i+1\dagger}\right),
\end{align}
\label{UpperBand}
\end{subequations}
leading to
\begin{equation}
\hat{H}_{0}(J_2=J_3)\left(\hat{W}_{+,i}^{1/2\dagger}\ket{0}\right)=
2\sqrt{2}J_2\left(\hat{W}_{+,i}^{1/2\dagger}\ket{0}\right).    
\end{equation}
In \figref{FlatStates} (b) we sketch the different states created by the lower-band, zero-energy-band and upper-band operators, given by Eqs.~\eqref{LowerBand}, \eqref{ZeroBand} and \eqref{UpperBand} respectively. Since each of these states spans a plaquette formed by the sites $\{A_i,B_i,C_i,B_{i+1},C_{i+1}\}$, a diamond chain with $N_c$ unit cells has $N_c-1$ compact localized modes of each type. Note that, even though states localized at consecutive plaquettes share a $B$ and a $C$ site, the total overlap is always zero, that is, these states form an orthogonal basis. As discussed in \cite{diamondchain1,diamondchain2}, a direct consequence of the fact that the single-particle spectrum consists entirely of highly localized eigenmodes is the appearance of Aharonov-Bohm caging \citep{ABcageoriginal}, i.e., the confinement of non-interacting wavepackets in small regions of the lattice due to
quantum interference.

After examining the single-particle eigenstates of the system in the flat-band limit, we are now in a position to include the effect of the on-site interactions, described by \refeq{Hint_dc}. For concreteness, we will focus on the case of attractive interactions, $U_A,U_1,U_2<0$, and analyze their effect on the low-energy properties of the two-boson states. However, we note that the procedure that we present below for the lowest flat bands can be carried out in an analogous way for states belonging to higher bands and for attractive or repulsive interactions.

In the absence of interactions there are many degenerate ground states of energy $E=-4\sqrt{2}J_2$, which consist of the two bosons occupying any of the localized modes belonging to the lowest flat bands. However, the introduction of an attractive interaction changes this scenario. From \refeq{Hint_dc}, it is clear that whenever two or more bosons occupy the same site an interaction term arises. Therefore, $\hat{H}_{\text{int}}$ yields non-zero matrix elements between two-boson states formed by localized modes which overlap to some extent. We concentrate on the case in which the interaction strength is much smaller than the size of the band gap, $|U_A|,|U_1|,|U_2|\ll 2\sqrt{2}J_2$, in such a way that the bands do not mix and the low-energy properties of the system can be examined by projecting the total Hamiltonian \eqref{HamTotal} to the lowest flat bands. If the lattice would host edge states, the weakly interacting condition would be slightly more stringent and would read $|U_A|,|U_1|,|U_2|\ll 4\sqrt{2}J_2-2(1+\sqrt{2})J_2\approx 0.8 J_2$.

Due to the effect of the on-site interactions, the total set of states with the two bosons in the lowest single-particle energy bands, which we denote as $\mathcal{H}$, can be divided into the following 4 subspaces according to the values of the self-energy:

\begin{itemize}
\item $\mathcal{H}_1\equiv\left\{\hat{W}_{-,i}^{n\dagger}\hat{W}_{-,k}^{m\dagger}\ket{0}\right\}$, with $n,m=1,2$; $i=1,...,N_c-1$ and $|k-i|\ge 2$.

This is the subspace of states where the two bosons occupy localized modes separated by two unit cells or more. In these configurations there is no overlap between the two particles, and therefore the interaction energy is zero,
\begin{equation}
\brakett{0}{\hat{W}_{-,i}^{n}\hat{W}_{-,k}^{m}\hat{H}_{\text{int}}\hat{W}_{-,i}^{n\dagger}\hat{W}_{-,k}^{m\dagger}}{0}=0.
\label{selfenergyH1}
\end{equation}
\item $\mathcal{H}_2\equiv\left\{\hat{W}_{-,i}^{n\dagger}\hat{W}_{-,i+1}^{m\dagger}\ket{0}\right\}$, with $n,m=1,2$ and $i=1,...,N_c-2$.

This is the set of states in which the localized modes of the two atoms are localized in two consecutive unit cells, in such a way that they share the sites $B_{i+1}$ and $C_{i+1}$. Using Eqs.~\eqref{LowerBand}, we find that their self-energy is
\begin{equation}
\brakett{0}{\hat{W}_{-,i}^{n}\hat{W}_{-,i+1}^{m}\hat{H}_{\text{int}}\hat{W}_{-,i}^{n\dagger}\hat{W}_{-,i+1}^{m\dagger}}{0}=
\left\{
	       \begin{array}{ll}
		 \frac{U_1}{32} \text{ if $i$ is even} \\
		 \\
		 \frac{U_2}{32} \text{ if $i$ is odd}\\
	       \end{array}
	       \right.
\label{selfenergyH2}
\end{equation}
\item $\mathcal{H}_3\equiv\left\{\frac{1}{\sqrt{2}}\hat{W}_{-,i}^{n\dagger}\hat{W}_{-,i}^{n\dagger}\ket{0}\right\}$, with $n=1,2$ and $i=1,...,N_c-2$.

This subspace is formed by the two-boson states in which both bosons occupy the same single-particle state. Since the localized modes overlap completely, there are interaction terms coming from the contributions of the  $A_{i},B_{i},C_{i},B_{i+1}$ and $C_{i+1}$ sites. Using again the expansions \eqref{LowerBand} we find that the self-energy of these states is
\begin{equation}
\brakett{0}{\frac{1}{\sqrt{2}}\hat{W}_{-,i}^{n}\hat{W}_{-,i}^{n}\hat{H}_{\text{int}}\frac{1}{\sqrt{2}}\hat{W}_{-,i}^{n\dagger}\hat{W}_{-,i}^{n\dagger}}{0}=
\frac{U_A}{4}+\frac{3\left(U_1+U_2\right)}{64}.
\label{selfenergyH3}
\end{equation}
\item $\mathcal{H}_4\equiv\left\{\hat{W}_{-,i}^{1\dagger}\hat{W}_{-,i}^{2\dagger}\ket{0}\right\}$, with $i=1,...,N_c-2$.

This subspace is formed by two-boson states in which the two atoms are localized in the same plaquette but occupy orthogonal states, each belonging to one of the two degenerate bands. As in the case of the $\mathcal{H}_3$ subspace, the two particles share 5 different sites. The self-energy of these states is

\begin{equation}
\brakett{0}{\hat{W}_{-,i}^{1}\hat{W}_{-,i}^{2}\hat{H}_{\text{int}}\hat{W}_{-,i}^{1\dagger}\hat{W}_{-,i}^{2\dagger}}{0}=
\frac{U_A}{2}+\frac{3\left(U_1+U_2\right)}{32}.
\label{selfenergyH4}
\end{equation}
\end{itemize}
Comparing Eqs.~\eqref{selfenergyH1}-\eqref{selfenergyH4}, we find that the subspace in which the states have a lowest self-energy is $\mathcal{H}_4$, followed by $\mathcal{H}_3$. In the next section, we derive effective models for these subspaces and we discuss their topological properties. Since the states of $\mathcal{H}_2$ have double occupancy only on the two common sites of two consecutive plaquettes, the spectrum of this subspace consists of flat bands formed by combinations of states localized in two neighbouring plaquettes.
\section{Effective models for the lowest-energy subspaces}
\label{EffectiveModels}
We can obtain effective models for the $\mathcal{H}_3$ and $\mathcal{H}_4$ subspaces by projecting the Hamiltonian \eqref{HamTotal} into the two lowest flat bands. In order to do so, we invert the relations of Eqs.~\eqref{LowerBand}, \eqref{ZeroBand} and \eqref{UpperBand}, and remove from the resulting expressions the contributions from higher bands \cite{FlatBands4}. This procedure yields the following non-orthonormal set of restricted OAM operators
\begin{subequations}
\begin{align}
\hat{\bar{a}}_{+}^{i\dagger}&=-\frac{1}{\sqrt{2}}\hat{W}_{-,i}^{1\dagger},\\
\hat{\bar{a}}_{-}^{i\dagger}&=-\frac{1}{\sqrt{2}}\hat{W}_{-,i}^{2\dagger},\\
\hat{\bar{b}}_{+}^{i\dagger}&=\frac{1}{4}\left(
\hat{W}_{-,i-1}^{1\dagger}+\hat{W}_{-,i-1}^{2\dagger}
+\hat{W}_{-,i}^{1\dagger}-\hat{W}_{-,i}^{2\dagger}\right),\\
\hat{\bar{b}}_{-}^{i\dagger}&=\frac{1}{4}\left(
\hat{W}_{-,i-1}^{1\dagger}+\hat{W}_{-,i-1}^{2\dagger}
-\hat{W}_{-,i}^{1\dagger}+\hat{W}_{-,i}^{2\dagger}\right),\\
\hat{\bar{c}}_{+}^{i\dagger}&=\frac{1}{4}\left(
\hat{W}_{-,i-1}^{1\dagger}-\hat{W}_{-,i-1}^{2\dagger}
+\hat{W}_{-,i}^{1\dagger}+\hat{W}_{-,i}^{2\dagger}\right),\\
\hat{\bar{c}}_{-}^{i\dagger}&=\frac{1}{4}\left(
-\hat{W}_{-,i-1}^{1\dagger}+\hat{W}_{-,i-1}^{2\dagger}
+\hat{W}_{-,i}^{1\dagger}+\hat{W}_{-,i}^{2\dagger}\right).
\end{align}
\label{expansionOAM_proj}
\end{subequations}
Inserting in \refeq{HamTotal} the projected OAM operators \eqref{expansionOAM_proj} instead of the original ones, we obtain a Hamiltonian $\hat{\bar{H}}$ expressed in the basis of the single-particle localized eigenstates and restricted to the two degenerate flat bands of lowest energy. In this basis, the non-interacting part of the projected Hamiltonian is diagonal and reads
\begin{equation}
\hat{\bar{H}}_{\text{kin}}=-2\sqrt{2}J_2\sum_{i=1}^{N_c-1}
\hat{W}_{-,i}^{1\dagger}\hat{W}_{-,i}^{1}+\hat{W}_{-,i}^{2\dagger}\hat{W}_{-,i}^{2}.
\label{HkinProj}
\end{equation}
The interacting part of the projected Hamiltonian, $\hat{\bar{H}}_{\text{int}}$, contains products of the type $\hat{\bar{j}}_{\alpha}^{i\dagger}\hat{\bar{j}}_{\alpha}^{i\dagger}
\hat{\bar{j}}_{\alpha}^{i}\hat{\bar{j}}_{\alpha}^{i}$ and $\hat{\bar{j}}_{\alpha}^{i\dagger}\hat{\bar{j}}_{\alpha}^{i\dagger}
\hat{\bar{j}}_{-\alpha}^{i}\hat{\bar{j}}_{-\alpha}^{i}$, with $j=\{a,b,c\}$ and $\alpha=\pm$. Therefore, insertion of Eqs.~\eqref{expansionOAM_proj} yields terms that involve products of operators that belong to the $\mathcal{H}_2$, $\mathcal{H}_3$ and $\mathcal{H}_4$ subspaces. Nevertheless, it can be shown that all the terms that couple states belonging to different subspaces cancel out. This fact greatly simplifies calculations, allowing to derive a separate projected Hamiltonian for each of the subspaces without making any further approximations. Moreover, the decoupling of the different low-energy subspaces is advantegous with regards to addressing them separately in an experimental implementation. Next, we present the effective models that are obtained for the two lowest-energy subspaces, $\mathcal{H}_4$ and $\mathcal{H}_3$. 
\subsection{$\mathcal{H}_4$ subspace}
\begin{figure}[h!]
\centering
\includegraphics[width=\linewidth]{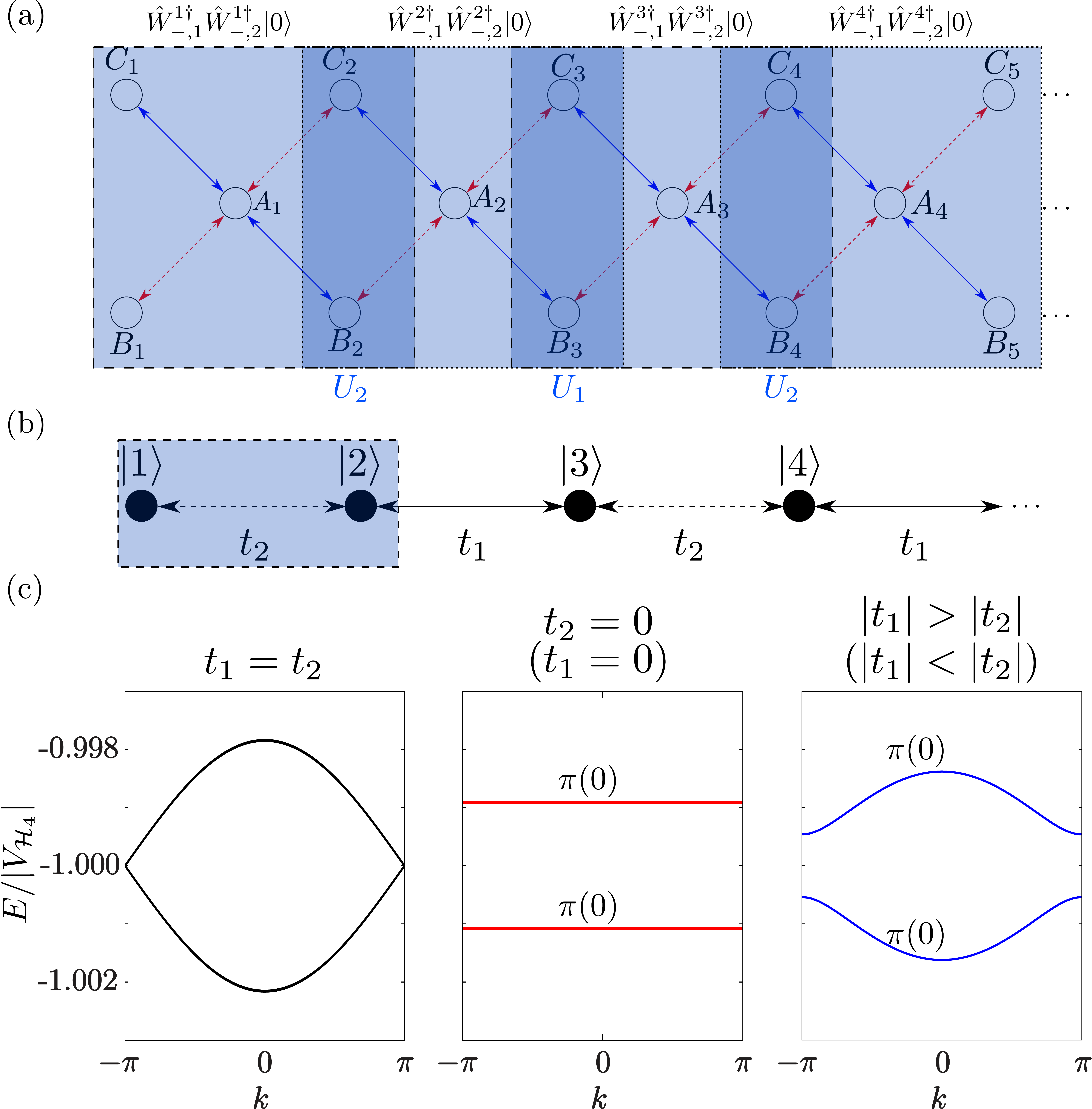}
\caption{Effective model for the $\mathcal{H}_4$ subspace. (a) Original states of $\mathcal{H}_4$ in the diamond chain. The blue squares framed in dashed lines indicate the plaquettes where the states are localized. In the sites with a darker blue shade, neighbouring states overlap and interact with a strength $U_1$ or $U_2$. (b) Effective SSH chain describing the $\mathcal{H}_4$ subspace. The blue shaded area indicates the unit cell, which is formed by two neighbouring plaquettes of the original diamond chain. (c) Band structure of the SSH chain for values of the interaction parameters $\{U_A/J_2=-0.2,U_1/J_2=-0.2,U_2/J_2=-0.2\}$ (left plot), $\{U_A/J_2=-0.2,U_1/J_2=-0.2(0),U_2/J_2=0(-0.2)\}$ (middle plot), and $\{U_A/J_2=-0.2,U_1/J_2=-0.2(-0.1),U_2/J_2=-0.1(-0.2)\}$ (right plot). In the middle and right plots, we also indicate the value of the Zak's phases of the bands for each case.}
\label{figure_H4Eff}
\end{figure}

The projected interaction Hamiltonian contains the following terms consisting of products of operators associated to the $\mathcal{H}_4$ subspace
\begin{align}
\hat{\bar{H}}_{\text{int}}^{\mathcal{H}_4}&=
\left(\frac{U_A}{2}+\frac{3\left(U_1+U_2\right)}{32}\right)
\sum_{i=1}^{N_c-1}\hat{W}_{-,i}^{1\dagger} \hat{W}_{-,i}^{2\dagger}\hat{W}_{-,i}^{1}\hat{W}_{-,i}^{2}\nonumber\\
&+\frac{U_1}{32}\sum_{i=1}^{[N_c/2]-2}\left(\hat{W}_{-,2i}^{1\dagger} \hat{W}_{-,2i}^{2\dagger}\hat{W}_{-,2i+1}^{1} \hat{W}_{-,2i+1}^{2}+\text{H.c.}\right)\nonumber\\
&+\frac{U_2}{32}\sum_{i=1}^{[N_c/2]-1}\left(\hat{W}_{-,2i-1}^{1\dagger} \hat{W}_{-,2i-1}^{2\dagger}\hat{W}_{-,2i}^{1} \hat{W}_{-,2i}^{2}+\text{H.c.}\right).
\label{HintProjH4}
\end{align}
The first term of \refeq{HintProjH4} is associated with the self-energy of the states of the $\mathcal{H}_4$ subspace, \refeq{selfenergyH4}, whereas the other two terms correspond to hoppings between two-boson states belonging to neighbouring plaquettes induced by the overlap between the localized modes, as shown in \figref{figure_H4Eff} (a). By mapping the two-body states of $\mathcal{H}_4$ into single particle states according to the definition $\ket{i}\equiv\hat{W}_{-,i}^{1\dagger} \hat{W}_{-,i}^{2\dagger}\ket{0}$, we can compute all the matrix elements of \refeqs{HkinProj} and \eqref{HintProjH4} over the two-boson states of  $\mathcal{H}_4$ and write an effective single-particle tight-binding model for this subspace,
\begin{align}
H_{\text{eff}}^{\mathcal{H}_4}
&=V_{\mathcal{H}_4}\sum_{i=1}^{N_c-1}\ket{i}\bra{i}\nonumber\\
&+t_1\sum_{i=1}^{N_{c}/2-2}\left(\ket{2i}\bra{2i+1}+\text{H.c.}\right)\nonumber\\
&+t_2\sum_{i=1}^{N_{c}/2-1}\left(\ket{2i-1}\bra{2i}+\text{H.c.}\right),
\label{HeffH4}
\end{align}
where we have defined $t_1\equiv \frac{U_1}{32}$, $t_2\equiv \frac{U_2}{32}$ and ${V_{\mathcal{H}_4}\equiv \left(-4\sqrt{2}J_2+\frac{U_A}{2}+\frac{3(U_1+U_2)}{32}\right)}$. As illustrated in \figref{figure_H4Eff} (b), \refeq{HeffH4} describes a Su-Schrieffer-Heeger (SSH) chain \cite{SSHoriginal} with a unit cell formed by two sites that correspond to neighbouring plaquettes of the original diamond chain. The intra- and inter-cell hoppings of this chain are given by $t_2$ and $t_1$ respectively, and the energy of all sites is shifted by a uniform potential $V_{\mathcal{H}_4}$.
By Fourier-transforming the Hamiltonian \eqref{HeffH4}, we find that the two energy bands of this model are given by
\begin{subequations}
\begin{align}
&E_{\mathcal{H}_4}^1(k)=V_{\mathcal{H}_4}-\sqrt{t_1^2+t_2^2+2t_1t_2\cos k},\\
&E_{\mathcal{H}_4}^2(k)=V_{\mathcal{H}_4}+\sqrt{t_1^2+t_2^2+2t_1t_2\cos k}.
\end{align}
\label{bands_SSH_H4}
\end{subequations}
The shape of the band structure \eqref{bands_SSH_H4} is shown in \figref{figure_H4Eff} (c) for different values of the hopping parameters. As shown in the left plot, for the particular case $t_1=t_2$ the gap closes at $k=\pm \pi$. As can be seen in the middle plot, for $t_1=0$ or $t_2=0$ the bands are flat and separated by a gap. In the general case $t_1\neq t_2$, which is illustrated in the right plot, both bands are dispersive and the gap remains open. In the situations where there is an energy gap, the relative value of the hopping parameters determines the topological properties of the system. If $|t_1|<|t_2|$ (i.e., if $|U_1|<|U_2|$), the Zak's phases \cite{ZakPhase} of the bands are $\gamma_{\mathcal{H}_4}^1=\gamma_{\mathcal{H}_4}^2=0$ and the system is in a topologically trivial phase. On the other hand, if the values of the hoppings are such that $|t_1|>|t_2|$ (i.e., if $|U_1|>|U_2|$), the Zak's phases are $\gamma_{\mathcal{H}_4}^1=\gamma_{\mathcal{H}_4}^2=\pi$ and the system is in a topological phase. Thus, by introducing the different interaction strengths $U_1$ and $U_2$ at $B$ and $C$ sites of odd and even unit cells respectively, we are able to control the shape of the band structure and to render the $\mathcal{H}_4$ subspace topologically non-trivial. According to the bulk-boundary correspondence, in the topological phase we expect a chain with open boundaries to display two edge states of energy $V_{\mathcal{H}_4}$ in its spectrum. Remarkably, these edge states consist of bound pairs of bosons and are induced by the interplay between the strength of different on-site interactions in a system where the kinetic energy plays no role due to the fact that all the bands are flat.
\subsection{$\mathcal{H}_3$ subspace}
\begin{figure}[t!]
\centering
\includegraphics[width=\linewidth]{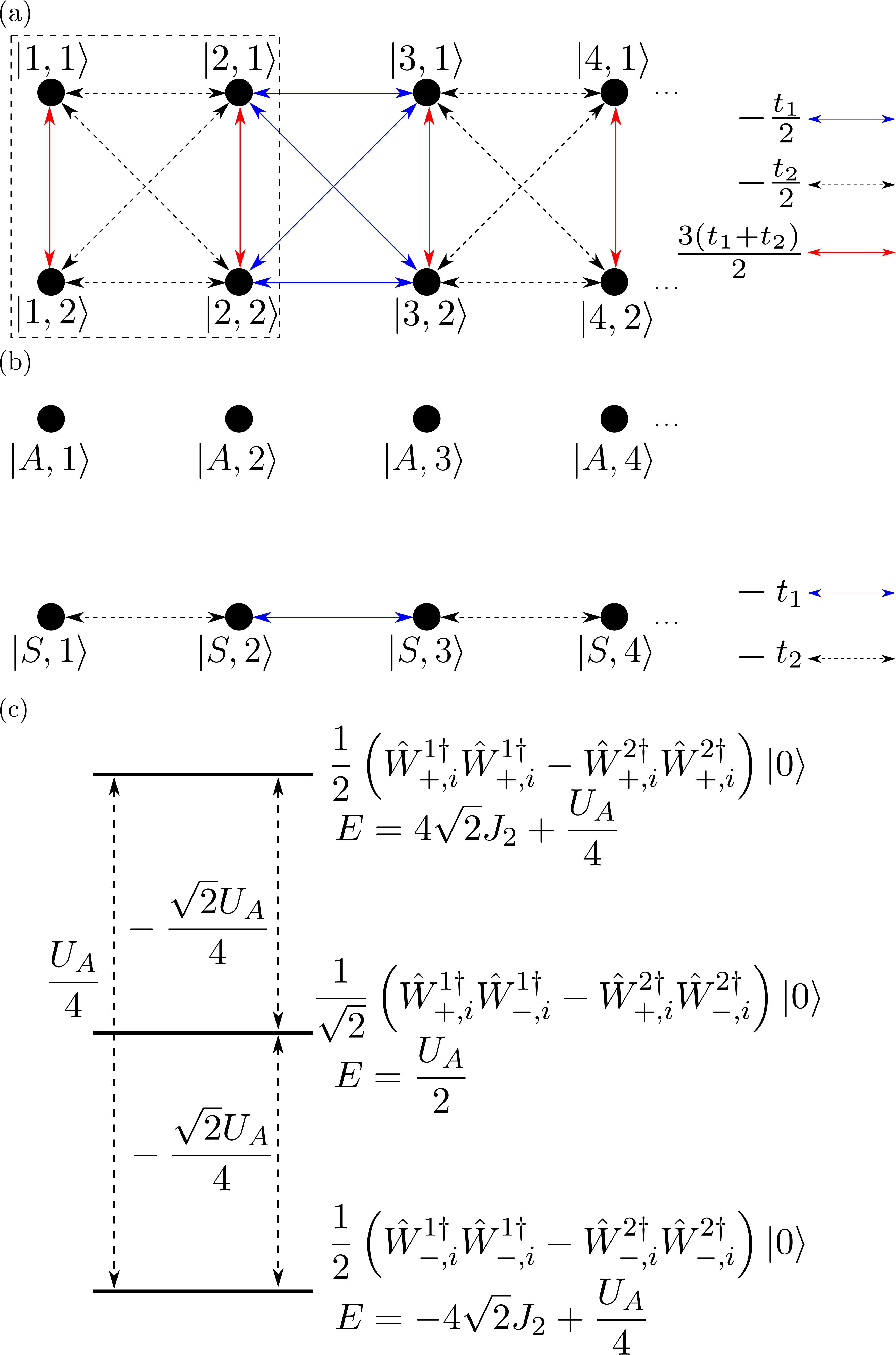}
\caption{(a) Effective Creutz ladder describing the $\mathcal{H}_3$ subspace. The area framed in dashed lines indicates the unit cell, which is formed by two neighbouring legs that are in turn composed of the two states of the original diamond chain localized in each plaquette. (b) Collection of isolated states (upper sketch) and SSH chain (lower sketch) formed respectively by the sets of anti-symmetric and symmetric combinations of states of each leg of the ladder. (c) Schematic representation of the closed system formed by the three doubly-localized flat-band states of each plaquette.}
\label{figure_H3Eff}
\end{figure}

The part of the interaction Hamiltonian projected to the two lowest flat bands containing products of operators associated to the $\mathcal{H}_3$ subspace reads
\begin{align}
&\bar{\hat{H}}_{\text{int}}^{\mathcal{H}_3}\nonumber\\
&=\left(\frac{U_A}{4}+\frac{3(U_1+U_2)}{64}\right)\sum_{i=1}^{N_c-1}\sum_{n=1,2}\frac{\hat{W}_{-,i}^{n\dagger} \hat{W}_{-,i}^{n\dagger}}{\sqrt{2}}\frac{\hat{W}_{-,i}^{n}\hat{W}_{-,i}^{n}}{\sqrt{2}}\nonumber\\
&+\frac{3\left(U_1+U_2\right)}{64}\sum_{i=1}^{N_c}\frac{\hat{W}_{-,i}^{1\dagger} \hat{W}_{-,i}^{1\dagger}}{\sqrt{2}}\frac{\hat{W}_{-,i}^{2} \hat{W}_{-,i}^{2}}{\sqrt{2}}+\text{h.c.}\nonumber\\
&-\frac{U_1}{64}\sum_{i=1}^{[N_c/2]-2}\sum_{n,m=1,2}\frac{\hat{W}_{-,2i}^{n\dagger} \hat{W}_{-,2i}^{n\dagger}}{\sqrt{2}}\frac{\hat{W}_{-,2i+1}^{m} \hat{W}_{-,2i+1}^{m}}{\sqrt{2}}+\text{H.c.}\nonumber\\
&-\frac{U_2}{64}\sum_{i=1}^{[N_c/2]-1}\sum_{n,m=1,2}\frac{\hat{W}_{-,2i-1}^{n\dagger} \hat{W}_{-,2i-1}^{n\dagger}}{\sqrt{2}}\frac{\hat{W}_{-,2i}^{m} \hat{W}_{-,2i}^{m}}{\sqrt{2}}+\text{H.c.}
\label{HintProjH3}
\end{align}
The first term of \refeq{HintProjH3} corresponds to the self-energy of the states belonging to $\mathcal{H}_3$, \refeq{selfenergyH3}, and the other ones can be regarded as pair-tunneling terms induced by the interactions at the sites where the single-particle localized modes interact. More specifically, the second term corresponds to a coupling between two-boson states localized at the same plaquette but belonging to different bands, and the third and fourth terms to hoppings between states, either in the same or different bands, localized in neighbouring plaquettes, respectively. By mapping the two-body states of $\mathcal{H}_3$ into single-particle states according to the definitions $\ket{i,1}\equiv\frac{1}{\sqrt{2}}\hat{W}_{-,i}^{1\dagger} \hat{W}_{-,i}^{1\dagger}\ket{0}$ and $\ket{i,2}\equiv\frac{1}{\sqrt{2}}\hat{W}_{-,i}^{2\dagger} \hat{W}_{-,i}^{2\dagger}\ket{0}$, we can combine all the non-zero matrix elements of the Hamiltonians \eqref{HintProjH3} and \eqref{HkinProj} over two-boson states of $\mathcal{H}_3$ and write an effective single-particle model for this subspace  
\begin{align}
\hat{H}_{\text{eff}}^{\mathcal{H}_3}&=V_{\mathcal{H}_3}\sum_{i=1}^{N_c}
\left(\ket{i,1}\bra{i,1}+\ket{i,2}\bra{i,2}\right)\nonumber\\
&+\frac{3\left(t_1+t_2\right)}{2}\sum_{i=1}^{N_c}\left(\ket{i,1}\bra{i,2}+\text{H.c.}\right)\nonumber\\
&-\frac{t_1}{2}\sum_{i=1}^{[N_c/2]-2}\sum_{n,m=1,2}\left(\ket{2i,n}\bra{2i+1,m}+\text{H.c.}\right)\nonumber\\
&-\frac{t_2}{2}\sum_{i=1}^{[N_c/2]-1}\sum_{n,m=1,2}\left(\ket{2i-1,n}\bra{2i,m}+\text{H.c.}\right),
\label{HeffH3}
\end{align}
where $V_{\mathcal{H}_3}\equiv\left(-4\sqrt{2}J_2+\frac{U_A}{4}+\frac{3(U_1+U_2)}{64}\right)$ and $t_1, t_2$ are defined in the same way as in \refeq{HeffH4}. As illustrated in \figref{figure_H3Eff} (a), \refeq{HeffH3} describes a Creutz ladder \cite{CreutzLadder} with a unit cell formed by two legs, each of which corresponds to the two states of $\mathcal{H}_3$ localized in each plaquette, $\ket{i,1}$ and $\ket{i,2}$. The intra-leg coupling is given by $\frac{3(t_1+t_2)}{2}$, and the inter-leg hoppings are $-\frac{t_1}{2}$ or $-\frac{t_2}{2}$ depending on the parity of the unit cell. Additionally, the energy of all sites is shifted by a uniform potential $V_{\mathcal{H}_3}$. By Fourier-transforming the Hamiltonian \eqref{HeffH3}, we find the following energy bands
\begin{subequations}
\begin{align}
&E_{\mathcal{H}_3}^1(k)=\frac{3(t_1+t_2)}{2}+V_{\mathcal{H}_3}-\sqrt{t_1^2+t_2^2+2t_1t_2\cos k},\\
&E_{\mathcal{H}_3}^2(k)=\frac{3(t_1+t_2)}{2}+V_{\mathcal{H}_3}+\sqrt{t_1^2+t_2^2+2t_1t_2\cos k},\\
&E_{\mathcal{H}_3}^3(k)=E_{\mathcal{H}_3}^4(k)=-\frac{3(t_1+t_2)}{2}+V_{\mathcal{H}_3}=-4\sqrt{2}J_2+\frac{U_A}{4}.
\end{align}
\label{bands_Creutz_H3}
\end{subequations}
Additional insight into the properties of the effective Hamiltonian \eqref{HeffH3} can be gained by performing a basis rotation into the symmetric and anti-symmetric combinations of the two states forming each leg of the ladder, $\ket{S,i}=\frac{1}{\sqrt{2}}\left(\ket{i,1}+\ket{i,2}\right)$ and $\ket{A,i}=\frac{1}{\sqrt{2}}\left(\ket{i,1}-\ket{i,2}\right)$. As illustrated in \figref{figure_H3Eff} (b), in this basis the effective Hamiltonian of the $\mathcal{H}_3$ subspace gets decoupled into two differents terms containing only symmetric and anti-symmetric states
\begin{align}
\hat{H}_{\text{eff}}^{\mathcal{H}_3}&=\hat{H}_{\text{eff}}^{\mathcal{H}_3}(S)+\hat{H}_{\text{eff}}^{\mathcal{H}_3}(A);\\
\hat{H}_{\text{eff}}^{\mathcal{H}_3}(S)&=\left(V_{\mathcal{H}_3}+\frac{3(t_1+t_2)}{2}\right)\sum_{i=1}^{N_c-1}\ket{S,i}\bra{S,i}\nonumber\\
&-t_1\sum_{i=1}^{N_{c}/2-2}\left(\ket{S,2i}\bra{S,2i+1}+\text{H.c.}\right)\nonumber\\
&-t_2\sum_{i=1}^{N_{c}/2-1}\left(\ket{S,2i-1}\bra{S,2i}+\text{H.c.}\right),\label{HeffH3_sym}\\
\hat{H}_{\text{eff}}^{\mathcal{H}_3}(A)&=\left(V_{\mathcal{H}_3}-\frac{3(t_1+t_2)}{2}\right)\sum_{i=1}^{N_c-1}\ket{A,i}\bra{A,i}.
\end{align}
On the one hand, the term corresponding to the symmetric states, $\hat{H}_{\text{eff}}^{\mathcal{H}_3}(S)$, is formally equivalent to the SSH model effectively describing the $\mathcal{H}_4$ subspace, \refeq{HeffH4}, except for a difference in the constant energy offset and a global change of sign in the effective couplings. Accordingly, this part of the Hamiltonian yields the energy bands $E_{\mathcal{H}_3}^1(k)$ and $E_{\mathcal{H}_3}^2(k)$, which follow the same dispersion law as the ones from the the $\mathcal{H}_4$ subspace, \refeqs{bands_SSH_H4}, and share the same topological phases. On the other hand, the term corresponding to the anti-symmetric states, $\hat{H}_{\text{eff}}^{\mathcal{H}_3}(A)$, describes a set of decoupled states with a constant energy offset ${V_{\mathcal{H}_3}-\frac{3(t_1+t_2)}{2}=-4\sqrt{2}J_2+\frac{U_A}{4}}$. Therefore, the $\ket{A,i}$ states form the flat bands $E_{\mathcal{H}_3}^3(k)$ and $E_{\mathcal{H}_3}^4(k)$, which are topologically trivial. These flat-band states feature two simultaneous localization effects, due to the fact that they are localized states with finite weight on a set of two-boson basis states which are, themselves, combinations of single-particle localized states. Another interesting property of this subspace of $\mathcal{H}_3$ is that each doubly-localized state $\ket{A,i}=\frac{1}{2}\left(\hat{W}_{-,i}^{1\dagger} \hat{W}_{-,i}^{1\dagger}-\hat{W}_{-,i}^{2\dagger} \hat{W}_{-,i}^{2\dagger}\right)\ket{0}$ is only coupled to two doubly-localized states of an analogous form belonging to higher bands, $\frac{1}{\sqrt{2}}\left(\hat{W}_{+,i}^{1\dagger} \hat{W}_{-,i}^{1\dagger}-\hat{W}_{+,i}^{2\dagger} \hat{W}_{-,i}^{2\dagger}\right)\ket{0}$ and $\frac{1}{2}\left(\hat{W}_{+,i}^{1\dagger} \hat{W}_{+,i}^{1\dagger}-\hat{W}_{+,i}^{2\dagger} \hat{W}_{+,i}^{2\dagger}\right)\ket{0}$, through the terms of \refeq{Hint_dc} proportional $U_A$ alone. Thus, at each plaquette $i$ these states form a closed three-level system for any value of the interaction parameters. Expressing the two-boson flat-band states in terms of the original OAM states and computing the corresponding matrix elements with the total Hamiltonian \eqref{HamTotal}, we find that the three-state Hamiltonian describing each of these sets of decoupled states is given by
\begin{equation}
H_{3L}=\begin{pmatrix}
-4\sqrt{2}J_2+\frac{U_A}{4} & -\sqrt{2}\frac{U_A}{4} & \frac{U_A}{4}\\
-\sqrt{2}\frac{U_A}{4} & \frac{U_A}{2} & -\sqrt{2}\frac{U_A}{4}\\
\frac{U_A}{4} & -\sqrt{2}\frac{U_A}{4} & -4\sqrt{2}J_2+\frac{U_A}{4}\\
\end{pmatrix}.
\label{Ham3L}
\end{equation}
A sketch of the three-level system described by \refeq{Ham3L} is shown in \figref{figure_H3Eff} (c). As we will show in the next section, the existence of this collection of closed three-level systems gives rise to a special type of two-boson Aharonov-Bohm caging effect for any value of the on-site interaction strengths.
\section{Exact diagonalization results}
\label{EDsec}
\begin{figure*}
\centering
\includegraphics[width=\linewidth]{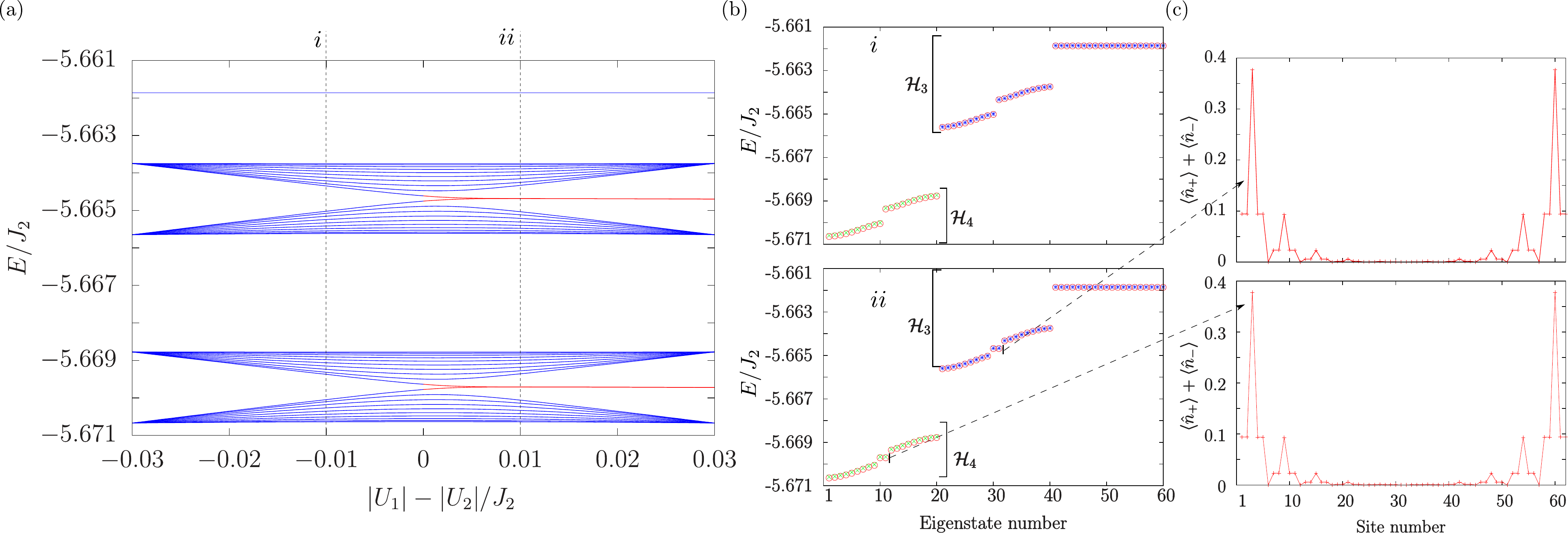}
\caption{(a) Low-energy sector of the exact diagonalization spectrum of a diamond chain formed by $N_c=21$ unit cells with open boundaries as a function of the difference between the interaction strengths at odd and even unit cells, $|U_1|-|U_2|$. Blue (red) curves representy bulk (edge) states. The parameters of the system fulfill the relations $J_2=J_3$, $U_A=-0.02J_2, U_1+U_2=-0.03J_2$. (b) Exact diagonalization spectrum (red empty dots) for the $i$ and $ii$ cases given by the vertical dashed lines in (a), compared against the spectra of the SSH model describing the $\mathcal{H}_4$ subspace (green dots) and the Creutz ladder describing the $\mathcal{H}_3$ subspace (blue dots). The parameters of the system are $\{U_A/J_2=-0.02,U_1/J_2=-0.01,U_2/J_2=-0.02\}$ (upper plot) and $\{U_A/J_2=-0.02,U_1/J_2=-0.02,U_2/J_2=-0.01\}$ (lower plot) (c) Density profiles of the in-gap states that appear in the  exact diagonalization spectrum of the lower plot of (b). The sites have been assigned a number $j$ according to the correspondence $A_i=3i, B_i=3i-1, C_i=3i-2$, where $i$ labels the unit cell.}
\label{EDresults}
\end{figure*}
In this section, we present exact diagonalization results that support the analysis of the lowest-energy subspaces discussed above. We also explore numerically the effects of deviations from the weakly-interacting and flat-band limits. All the calculations have been performed considering a diamond chain formed by $N_c=21$ unit cells (and therefore $3N_c-1=62$ sites) and filled with $N=2$ bosons. We focus on the first $3(N_c-1)$ states of the spectrum, of which the $N_c-1$ states of lowest energy correspond to the $\mathcal{H}_4$ subspace, and the remaining $2(N_c-1)$ states to the $\mathcal{H}_3$ subspace. 

In \figref{EDresults} (a) we show the spectrum obtained by diagonalizing the full Hamiltonian of the system, \refeq{HamTotal}, as a function of the difference between the interaction strengths at the $B$ and $C$ sites of odd and even unit cells, $|U_1|-|U_2|$. The tunneling parameters are set in the flat-band limit, $J_2=J_3$, and the on-site interaction strengths are $U_A=-0.02J_2, U_1+U_2=-0.03J_2$, fulfilling the weakly-interacting condition $|U_A|,|U_1|,|U_2|\ll 2\sqrt{2}J_2$. As predicted by the analysis of the band structures of the effective models describing the $\mathcal{H}_4$ and $\mathcal{H}_3$ subspaces, for $|U_1|-|U_2|>0$ the system is in a topological phase characterized by the presence of in-gap edge states in the spectrum, which correspond to the red lines of \figref{EDresults} (a). In \figref{EDresults} (b) we compare the energy spectra obtained by exact diagonalization of the full Hamiltonian (red empty dots), the effective Hamiltonian of the $\mathcal{H}_4$ subspace, \refeq{HeffH4}, (green dots) and the effective Hamiltonian of the $\mathcal{H}_3$ subspace, \refeq{HeffH3}, (blue dots). The plot $(i)$ displays the spectra corresponding to the interaction parameters $\{U_A/J_2=-0.02,U_1/J_2=-0.01,U_2/J_2=-0.02\}$ (non-topological phase), while $(ii)$ displays the spectra corresponding to the parameters $\{U_A/J_2=-0.02,U_1/J_2=-0.02,U_2/J_2=-0.01\}$ (topological phase). In both cases we observe that the effective models fit very well with the results obtained by tackling the full system. In \figref{EDresults} (c) we plot the total density profiles (i.e., the sum of the populations of the two OAM states at each site) of the in-gap states of the $\mathcal{H}_4$ and $\mathcal{H}_3$ subspaces that appear in the exact two-boson spectrum of \figref{EDresults} (b) $(ii)$. The sites of the chain have been assigned a number according to the correspondence $A_i=3i, B_i=3i-1, C_i=3i-2$. As one expects because of their topological origin, these states are strongly localized at the edges of the diamond chain, with the population peaks occurring at the first and last $A$ sites of the diamond chain, $A_1$ and $A_{N_c-1}$. The localization in these specific sites is due to the fact that the compact localized modes corresponding to the lowest bands, \refeqs{LowerBand}, have four times more population on the $A$ sites than on the $B$ and $C$ sites. Accordingly, the edge states of \figref{EDresults}, which are formed of bound pairs of these modes, also concentrate their population on these sites. The similarity between the population distributions of the edge states corresponding to the two different subspaces is a consequence of the underlying equivalence between the SSH model describing the $\mathcal{H}_4$ subspace, \refeq{HeffH4}, and the term of the Creutz ladder describing $\mathcal{H}_3$ given by \eqref{HeffH3_sym}, which corresponds to the symmetric combinations of orbitals at each leg.
\subsection*{Two-boson Aharonov-Bohm caging}
\begin{figure}[t!]
\centering
\includegraphics[width=\linewidth]{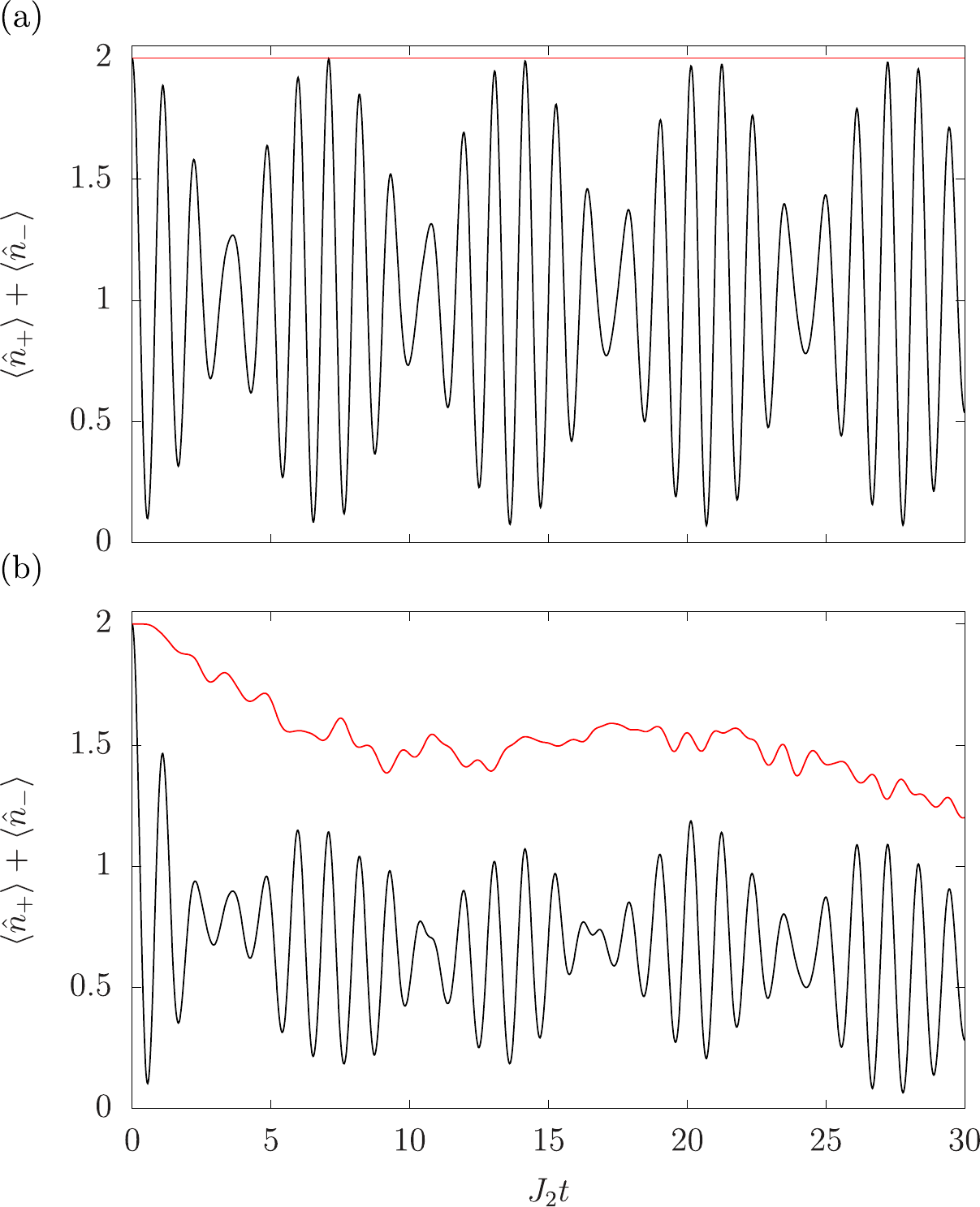}
\caption{Time evolution of the total population in site $A_4$ (black lines) and the sum over the cage formed by the sites $\{A_4,B_4,C_4,B_{5},C_{5}\}$ (red lines) in a chain formed by 8 unit cells and with onsite interactions $U_A=U_1=2U_2=-2J_2$. The initial state is $\frac{1}{2}\left[(\hat{a}_{+}^{4\dagger})^2-(\hat{a}_{-}^{4\dagger})^2\right]\ket{0}$ in (a) and $(\hat{a}_{+}^{4\dagger})^2\ket{0}$ in (b).}
\label{ABcaging}
\end{figure}

The impact of interactions on the Aharonov-Bohm caging effect has been studied in different scenarios \cite{ABinter2,ABinter1,ABinter2part1,ABinter2part2}. At the two-body level, it has been found that interactions break the cages and enable the spreading of the particles through the entire lattice \cite{ABinter2,ABinter2part1,ABinter2part2}. However, in the system considered here the OAM degree of freedom gives rise to robust Aharonov-Bohm caging for a particular instance of two-boson states. Specifically, this phenomenon stems from the existence at each plaquette of the chain of a three-level system formed by doubly-localized states (each constructed by a different combination of single-particle localized states of the upper and lower bands) described by the Hamiltonian \eqref{Ham3L}. Since each of these subsystems is decoupled from the rest of two-boson states, any combination of states of a given three-level system evolves coherently between the three different doubly-localized states. As an example, let us consider a state in which one boson is loaded into a symmetric combination of OAM states ($p_x$-like orbital) and the other in anti-symmetric combination ($p_y$-like orbital) at the site $A_i$. Such a state can be expressed in terms of doubly-localized two-boson states as  
\begin{align*}
&\left(\frac{\hat{a}_{+}^{i\dagger}+\hat{a}_{-}^{i\dagger}}{\sqrt{2}}\right)\left(\frac{\hat{a}_{+}^{i\dagger}-\hat{a}_{-}^{i\dagger}}{\sqrt{2}}\right)\ket{0}=\frac{1}{2}\left[(\hat{a}_{+}^{i\dagger})^2-(\hat{a}_{-}^{i\dagger})^2\right]\ket{0}=\nonumber\\
&\left[\frac{1}{2}\left(\frac{\hat{W}_{-,i}^{1\dagger} \hat{W}_{-,i}^{1\dagger}}{2}-\frac{\hat{W}_{-,i}^{2\dagger}\hat{W}_{-,i}^{2\dagger}}{2}\right)+\frac{1}{2}\left(\frac{\hat{W}_{+,i}^{1\dagger} \hat{W}_{+,i}^{1\dagger}}{2}-\frac{\hat{W}_{+,i}^{2\dagger}\hat{W}_{+,i}^{2\dagger}}{2}\right)\right.
\nonumber\\
&\left.-\frac{1}{\sqrt{2}}\left(\frac{\hat{W}_{+,i}^{1\dagger} \hat{W}_{-,i}^{1\dagger}}{\sqrt{2}}-\frac{\hat{W}_{+,i}^{2\dagger}\hat{W}_{-,i}^{2\dagger}}{\sqrt{2}}\right)\right]\ket{0}.
\end{align*}
Therefore, if one prepares the system in this initial state, its time evolution takes place within the three-level system formed by the doubly-localized flat-band states of the plaquette $i$. Thus, the total population of the time-evolved state remains inside the cage formed by the sites $\{A_i,B_i,C_i,B_{i+1},C_{i+1}\}$ regardless of the strength of the onsite interactions. This fact is illustrated in \figref{ABcaging} (a), where we plot, for a chain of $N_c=8$ unit cells and with interaction parameters $U_A=U_1=2U_2=-2J_2$, the time evolution of the total population in site $A_4$ (black line) and in the corresponding Aharonov-Bohm cage (red line) after injecting the initial state $\frac{1}{2}\left[(\hat{a}_{+}^{4\dagger})^2-(\hat{a}_{-}^{4\dagger})^2\right]\ket{0}$. In contrast, in \figref{ABcaging} (b), where we plot the same quantities for the initial state $(\hat{a}_{+}^{4\dagger})^2\ket{0}$, we observe a population leakage from the cage due to the coupling to neighbouring plaquettes induced by the onsite interactions. Nevertheless, since this state has a $1/\sqrt{2}$ projection to the perfectly caged state, there is a an upper bound of 1 boson to the total population that can escape the Aharonov-Bohm cage.   
\subsection*{Deviations from the weakly-interacting regime}
\begin{figure}[h!]
\centering
\includegraphics[width=\linewidth]{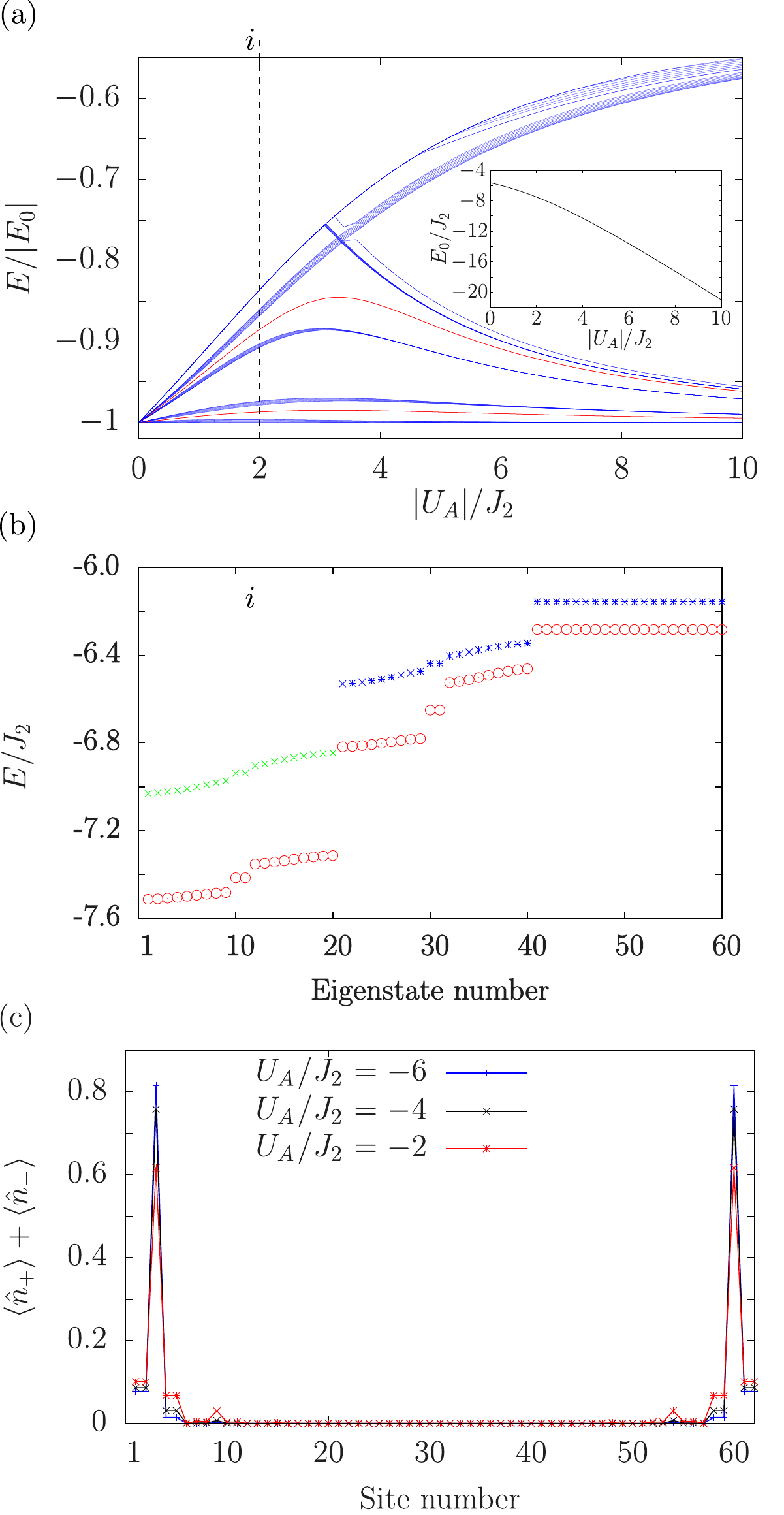}
\caption{(a) Low-energy sector of the exact diagonalization spectrum of a diamond chain formed $N_c=21$ unit cells with open boundaries as a function of the interaction strength at odd and even unit cells at the $A$ sites. Blue (red) curves representy bulk (edge) states. The inset shows the dependence on $|U_A|$ of the ground-state energy $E_0$, which sets the energy scale for each point of the main plot. The parameters of the system fulfill the relations $J_2=J_3$, $U_A=U_1=2U_2$, with $U_A<0$.  (b) Exact diagonalization spectrum (red empty dots), compared against the spectra of the SSH model describing the $\mathcal{H}_4$ subspace (green dots) and the Creutz ladder describing the $\mathcal{H}_3$ subspace (blue dots) for the interaction parameters $U_A=U_1=2U_2=-2J_2$. (c) Density profile of the lowest-energy edge states, belonging to the $\mathcal{H}_4$ subspace, for different values of $U_A$.}
\label{StrongInt}
\end{figure}

By means of exact diagonalization calculations, we can examine to which extent the effective models reproduce the features of the low-energy sector of the full spectrum when the interactions are strong enough to introduce couplings with higher bands. In \figref{StrongInt} (a) we show the energy spectrum for $J_2=J_3$ as a function of $|U_A|$ (with $U_A<0$). The interaction strengths fulfill the relation $U_A=U_1=2U_2$, for which the system is in the topological phase in the weakly-interacting regime. In the main plot, the energies are expressed in units of the ground state energy $E_0$, whose dependence on $|U_A|$ is shown on the inset. We observe that the edge states predicted by the effective models (signalled with red lines) remain in the middle of the energy gaps even for values of $|U_A|$ significantly larger than the single-particle energy gap, $2\sqrt{2}J_2$. Therefore, the topological properties of the system remain unaltered throughout the considered interaction strength sweep. For $|U_A|\gtrsim 3$, the highest energy states displayed in the main plot start to couple significantly with other states belonging to higher bands, giving rise to multiple energy crossings. In \figref{StrongInt} (b) we show, for $U_A=U_1=2U_2=-2J_2$, a comparison between the exact two-boson spectrum (red empty dots) and the effective models of the $\mathcal{H}_4$ (green dots) and $\mathcal{H}_3$ (blue dots) subspaces. Due to the mixing with higher bands, the energies obtained by exact diagonalization are shifted with respect to the ones predicted by the effective models. For the doubly localized flat-band states, the exact energies can be computed as the lowest eigenvalue of the three-level Hamiltonian describing the coupling to higher energetic two-boson flat bands, \refeq{Ham3L}.  

Although quantitatively the effective models do not agree perfectly with the exact diagonalization results away from the limit of small interactions, they still provide a good qualitative description of the lowest energetic two-boson states. In particular, in the exact spectrum of \figref{StrongInt} (b), the $\mathcal{H}_4$ and $\mathcal{H}_3$ subspaces are well separated in energy, the bands of each subspace maintain the shape predicted by the corresponding effective models, and, since the system is in the topological phase, there are in-gap states. In \figref{StrongInt} (c) we plot the density profiles of the lowest energetic edge states that appear in \figref{StrongInt} (a) for $|U_A|/J_2=2,4,6$. We observe that the localization of the states at the edge of the chain is more pronounced for higher interaction strengths, confirming the fact that the coupling to higher bands does not affect the topological properties of the system. From an experimental point of view, the sharpening of the localization of the two-boson edge states with the increase of the interaction strengths is advantageous for their detection, since it ensures that these states are recognizable for values of the interaction strength ranging from a small fraction of the tunneling energy to $|U|/J_2 \sim 10$.
\subsection*{Deviations from the flat-band limit}
\begin{figure}[h!]
\centering
\includegraphics[width=\linewidth]{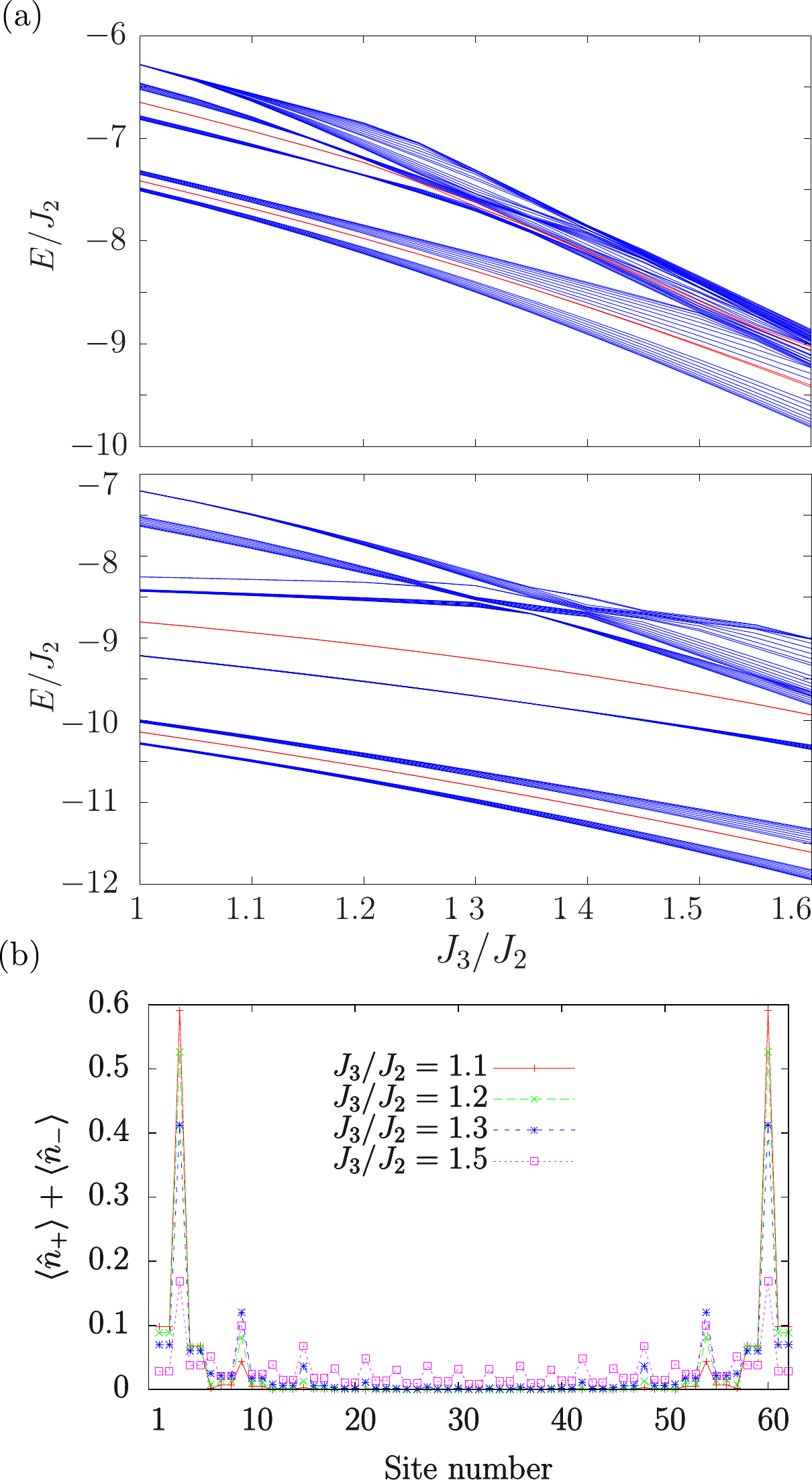}
\caption{(a) Low-energy sector of the exact diagonalization spectrum of a diamond chain formed $N_c=21$ unit cells with open boundaries as a function of the ratio between the two tunneling amplitudes between OAM states, $J_3/J_2$. Blue (red) curves representy bulk (edge) states. The parameters of the system fulfill the relations $U_A=U_1=2U_2=-2J_2$ (upper plot) and $U_A=U_1=2U_2=-4J_2$ (lower plot). (b) Density profile of the lowest-energy edge states, belonging to the $\mathcal{H}_4$ subspace, for $U_A=U_1=2U_2=-2J_2$ and different values of $J_3/J_2$.}
\label{NoFlat}
\end{figure}
So far, we have assumed that the system is in the flat-band limit, which occurs for $J_2=J_3$. However, as discussed in \cite{diamondchain1,diamondchain2}, in a real system $J_3$ is always slightly larger than $J_2$ even for large values of the inter-site separation $d$. Thus, it is relevant to examine the effect of deviations from the $J_2=J_3$ limit on the low-energy properties of the two-boson spectrum. For $J_2\neq J_3$, the non-interacting part of the projected Hamiltonian reads as
\begin{align}
\hat{\bar{H}}_{0}=&-\sqrt{2}(J_2+J_3)\sum_{i=1}^{N_c-1}
\hat{W}_{-,i}^{1\dagger}\hat{W}_{-,i}^{1}+\hat{W}_{-,i}^{2\dagger}\hat{W}_{-,i}^{2}\nonumber\\
&+\frac{(J_3-J_2)}{\sqrt{2}}\sum_{i=1}^{N_c-1}
\left(\hat{W}_{-,i}^{1\dagger}\hat{W}_{-,i+1}^{1}+\hat{W}_{-,i}^{2\dagger}\hat{W}_{-,i+1}^{2}+
\text{H.c.}\right).
\label{HkinProj2}
\end{align}
Due to the second term of \refeq{HkinProj2}, which corresponds to single-particle hoppings between adjacent localized eigenmodes, the flat bands become dispersive. Note that the set of all localized states forms a complete orthonormal basis, and one can always choose this basis to represent the Hamiltonian, regardless of the parameters. What \refeq{HkinProj2} shows is that, when $J_2\neq J_3$, this basis is no longer simultaneously the eigenbasis of the model, and it is no longer possible to derive effective single-particle models for the different subspaces of two-boson states. However, we can perform exact diagonalization calculations over the full Hamiltonian \eqref{HamTotal} to examine numerically to which extent the features of the system that we observe in the flat-band limit survive. In \figref{NoFlat} (a) we plot the energy spectrum of the system as a function of the $J_3/J_2$ ratio for $U_A=U_1=2U_2=-2J_2$ (upper plot) and $U_A=U_1=2U_2=-4J_2$ (lower plot). In the upper plot we observe that for $J_3/J_2\lesssim 1.2$ the separation between $\mathcal{H}_4$ and $\mathcal{H}_3$ is still clear and each of the subspaces preserves its in-gap states, signalled with red lines. As the $J_3/J_2$ ratio is increased, the energy separation between the two subspaces is lost and the edge states of $\mathcal{H}_3$ merge into the bulk. In the lower plot of \figref{NoFlat} (a), which corresponds to higher values of $|U|$, these effects are less pronounced and the edge states of both subspaces remain inside the energy gaps. Therefore, the topological two-boson states are more robust against deviations from the flat-band limit for stronger onsite interactions. In \figref{NoFlat} (b) we plot the total density profiles of the lowest-energy edge states for $U_A=U_1=2U_2=-2J_2$ and several values of the $J_3/J_2$ ratio, observing longer decays into the bulk as it is increased. From this brief analysis of the effects of deviations from the flat-band limit, we can conclude that for sufficiently low values of the difference between the $J_3$ and $J_2$ couplings (compared to the interaction strength $|U|$), the main characteristics of the low-energy sector of the spectrum, i.e., the separation between the $\mathcal{H}_4$ and $\mathcal{H}_3$ subspaces and the presence of topological edge states in the $|U_1|>|U_2|$ regime, are preserved. Thus, in a real experimental implementation, for which the $J_3/J_2$ ratio takes values $1.05\lesssim J_3/J_2<2$ \cite{diamondchain1,diamondchain2}, the main features of the flat-band limit could in principle be observed.
\section{Experimental considerations}
\label{Experiment}
A diamond-chain filled with ultracold atoms in OAM $l=1$ states could be realized in different ways. On the one hand, one could construct a lattice made up of ring potentials. In these trapping geometries, which have been experimentally realized over the last years by means of a variety of techniques such as optically plugged magnetic traps \cite{ring1}, static Laguerre-Gauss beams \cite{ring2}, painting \cite{ring3,ring4} and time-averaged potentials \cite{ring5,ring6,ringTaver,TAAP,Sagnac3} or conical refraction \cite{ring7}, OAM can be transferred to the atoms using light beams \cite{persistent2,OAMLightAtoms}, through temperature quenches \cite{QuenchSupercurrent} or by rotating a weak link \cite{squid1,squid2}. In order to create the desired geometrical arrangement of ring traps, one could adapt existing schemes for engineering arbitrary potential landscapes based on time-averaged adiabatic potentials or digital micro-mirror devices \cite{DMDBaker}, which have already been used to implement a double ring trap \cite{squid6b}. Alternatively, since the OAM $l=1$ states are equivalent to the $p_x$ and $p_y$ orbitals of the first excited Bloch band of a deep optical lattice \cite{HigherOrbital1}, one could implement the considered model by creating an optical lattice with a diamond-chain shape and exciting the atoms to the $p-$ band \cite{HigherOrbital2,HigherOrbital3,HigherOrbital4}. In such systems, the two-body bound states discussed in the previous sections could be imaged making use of high-resolution quantum gas microscopes \cite{QGasMicro1,QGasMicro2,QGasMicro3,QGasMicro4,QGasMicro5,QGasMicro6,DoublonHarperHofstadter}.

Once the atoms are loaded into the OAM $l=1$ states, loss of population could be induced by two-body collisions taking one atom to the ground state and the other to an OAM $l=2$ state. However, this effect could be greatly suppressed by tuning the system in the weakly-interacting regime, which, as discussed in the previous sections, is well suited for observing the predicted properties of the OAM $l=1$ bound states. Furthermore, in ring potentials the anharmonicity between the energies of the different OAM states ensures that these collisional processes are off-resonant, enhancing the stability of the interacting OAM $l=1$ states \cite{spinmodel}.
\section{Conclusions}       
\label{Conclusions}
We have proposed a mechanism based on the manipulation of interactions to obtain two-body topological states in systems with single-particle flat bands. As an example, we have studied a diamond-chain optical lattice filled with ultracold bosons in OAM $l=1$ states, wherein a proper adjustment of the tunneling parameters may lead to a band structure consisting of three two-fold degenerate flat bands formed by highly-localized eigenstates. In order to gain analytical insight into the properties of the two-boson states, we have first focused on the regime of small attractive interactions compared with the gap between the flat bands. By projecting the Hamiltonian into the lowest bands, we have analyzed the lowest-energy sector of the spectrum, which is composed of bound pairs of atoms occupying highly localized eigenmodes. Although the single-particle transport is suppressed because of the flatness of the bands, these composite objects can move through the lattice due to interaction-induced couplings. We have found that the effective models describing this motion can be rendered topologically non-trivial by tuning separately the strength of the interactions on different sites of the chain, with the appearance of the corresponding two-boson protected edge states. Furthermore, we have identified doubly-localized flat-band states that give rise to robust Aharonov-Bohm cages for a particular set of two-boson states. 

By means of exact diagonalization calculations, we have benchmarked our analytical predictions and we have tested the robustness of the topological two-boson states against deviations from the weakly-interacting and flat-band limits. In the former case, we have found that interaction strengths capable of introducing mixing between the bands do not destroy the topological phases and actually enhance the localization of the edge states at the ends of the chain. In the latter case, we have found that by going sufficiently away from the flat-band limit the edge states can merge into the bulk. However, this effect can be strongly mitigated by increasing the interaction strength, in such a way that the growth of the gaps between the two-boson bands outgains the curvature of the single-particle bands. Although in an actual experimental implementation the completely flat-band limit can not be reached, by introducing moderate interaction strengths the two-boson topological states should be observable in a feasible range of parameters. The procedure presented in this work could be readily adapted to study similar topological states of bound particles in other models featuring flat bands in one or higher dimensions, expanding the toolbox for the engineering of few-body topological systems.   
\acknowledgements
We thank Alexandre Dauphin for useful discussions. G.P., J.M., and V.A. gratefully acknowledge financial support  from  the  Ministerio  de  Econom\'ia  y  Competitividad, MINECO, (FIS2017-86530-P),  from the Generalitat de Catalunya (SGR2017-1646), and from the European Union Regional Development Fund within the ERDF Operational Program of Catalunya (project QUASICAT/QuantumCat). G.P. acknowledges  financial  support  from  MINECO  through  Grant  No. BES-2015-073772 and a travel grant from the COST Action CA16221. A.M.M. and R.G.D. developed their work within the scope of the Portuguese Institute for Nanostructures, Nanomodelling  and  Nanofabrication (i3N) project UIDB/50025/2020 $\&$ UIDP/50025/2020, and acknowledge funding from FCT - Portuguese Foundation for Science and Technology through the project PTDC/FIS-MAC/29291/2017. A.M.M.  acknowledges  financial  support from the FCT through the work contract CDL-CTTRI-147-ARH/2018. R.G.D. appreciates the support by the Beijing CSRC.

\bibliographystyle{apsrev4-1.bst}

\end{document}